\documentclass[9pt, conference]{IEEEtran}
\IEEEoverridecommandlockouts
\usepackage{cite}
\usepackage{amsmath,amssymb,amsfonts}
\usepackage{algorithmic}
\usepackage{graphicx}
\usepackage{textcomp}
\usepackage{xcolor}
\usepackage{booktabs}
\usepackage{tabularx}
\usepackage{array}
\usepackage{multirow}
\usepackage{colortbl}
\usepackage{subcaption}
\usepackage{url}

\definecolor{mastbg}{RGB}{221,235,247}

\newcommand{\mastcell}[1]{\cellcolor{mastbg}\textbf{#1}}

\def\BibTeX{{\rm B\kern-.05em{\sc i\kern-.025em b}\kern-.08em
    T\kern-.1667em\lower.7ex\hbox{E}\kern-.125emX}}
\begin{document}

\title{MicroAssert: Microarchitecture-Aware Structured Assertion Generation for CPU Verification
\thanks{* These authors contributed equally to this work.}
\thanks{\textdagger ~Corresponding authors.}
\thanks{Hongqin Lyu: lvhongqin24@mails.ucas.ac.cn; Yonghao Wang: wangyonghao@link.cuhk.edu.hk.}
}

\author{
\IEEEauthorblockN{
Hongqin Lyu\textsuperscript{1,2,*},
Yonghao Wang\textsuperscript{3,*}, 
Zhiteng Chao\textsuperscript{1}, 
Wenxing Li\textsuperscript{1},
Tiancheng Wang\textsuperscript{1},
Evangeline F.Y. Young\textsuperscript{3,\textdagger}
and
Huawei Li\textsuperscript{1,2,\textdagger}}
\IEEEauthorblockA{\textsuperscript{1}State Key Lab of Processors, Institute of Computing Technology, CAS;}
\IEEEauthorblockA{\textsuperscript{2}University of Chinese Academy of Sciences; \textsuperscript{3}The Chinese University of Hong Kong}
}

\maketitle

\begin{abstract}
 Automatic assertion generation reduces manual verification effort, yet existing approaches struggle to capture microarchitectural behaviors and cross-region interactions in CPUs. We present MicroAssert, a microarchitecture-aware framework that identifies functional regions directly from RTL and generates local and composite assertions through structured contracts and deterministic verification primitives. Coverage-guided feedback further improves verification completeness by progressively expanding proof-relevant RTL coverage. Evaluated on five open-source RISC-V CPUs against three existing assertion generation approaches, MicroAssert achieves 100\% syntax correctness and complete branch, statement, and toggle coverage, with an average proven rate of 74.82\% and Proof Core coverage of 90.96\%. Mutation testing further demonstrates its advantage, achieving an average error detection rate of 28.73\%, 1.99×–3.51× higher than existing methods. These results demonstrate the effectiveness of microarchitecture-aware assertion generation. 
\end{abstract}


\section{Introduction}
Functional verification is essential in integrated circuit (IC) design, where verification engineers check whether the designers' register-transfer level (RTL) code meets the architectural specifications. Assertion-based verification (ABV) is widely adopted in RTL design due to its ability to enhance observability and reduce simulation debugging time by up to 50\% \cite{b1}. In particular, high-quality SystemVerilog assertions (SVAs) for property verification (PV) are critical within ABV methodologies \cite{b2}, as they accurately reflect both high-level design intent and low-level RTL details. As design complexity continues to increase, developing efficient methods for SVA generation has become an urgent necessity.

Traditional assertion generation approaches include schema-based methods that instantiate predefined templates \cite{b3}, as well as waveform- or trace-mining techniques such as HARM \cite{b4} and GoldMine \cite{b5}. While effective for certain patterns, these methods often produce semantically fragmented assertions and remain limited in handling complex design semantics. More recently, large language models (LLMs) have opened new avenues for assertion generation \cite{b7, b23, agile, coverassert}, leveraging strong semantic understanding to produce comprehensive, context-aware assertions from specifications and RTL code, thereby mitigating semantic fragmentation \cite{b8}.

However, existing LLM-based assertion generation methods are largely organized around RTL modules or local code contexts. Such module-centric approaches overlook the fact that RTL modules primarily reflect implementation structures rather than the microarchitectural behaviors governing CPU execution. Moreover, existing methods typically apply a unified generation strategy to different verification objectives, without considering the diverse functional semantics of different microarchitectural components. These limitations make CPU-level assertion generation particularly challenging, as complex control flows, state transitions, and component interactions introduce verification requirements that cannot be adequately addressed by conventional module-centric approaches. Consequently, effective CPU-level assertion generation requires identifying functionally meaningful microarchitectural regions beyond conventional RTL module boundaries and performing verification guided by their intended functional behaviors.

Identifying such microarchitectural regions from RTL is challenging. Functionally similar behaviors may be implemented using different hardware structures, while a single structure may participate in multiple functional behaviors. Consequently, meaningful microarchitectural boundaries cannot be directly inferred from RTL module boundaries and instead require analyzing the underlying hardware structures and their interactions.   This analysis must jointly capture data dependencies, control flows, sequential behaviors, and state-transition relationships to distinguish functionally coherent regions.

\begin{figure}[t]
    \centering
    \includegraphics[width=\columnwidth]{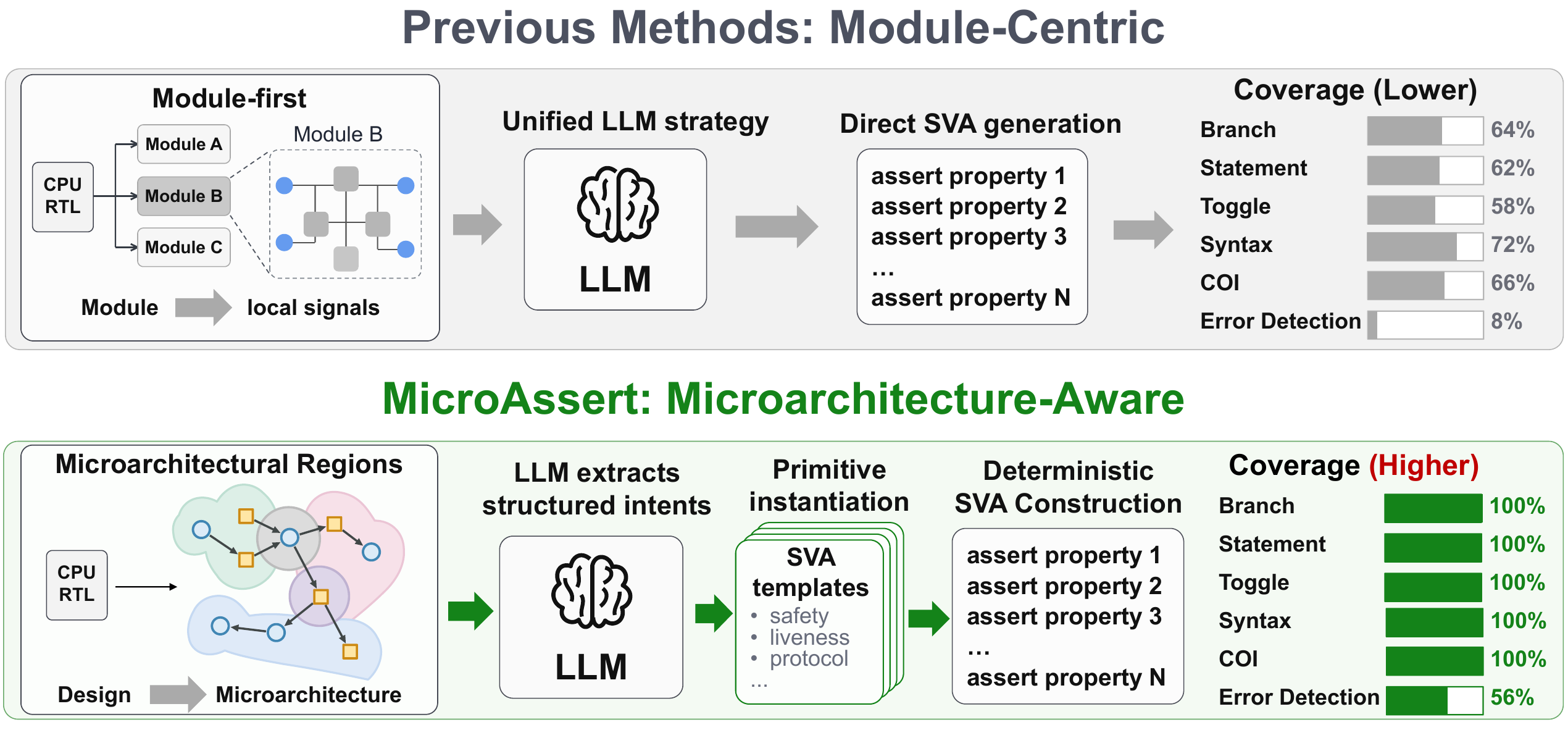}
    \caption{Unlike previous module-based approaches, MicroAssert identifies microarchitectural functional regions and generates assertions beyond RTL module boundaries.}
    \label{fig:motivation}
\end{figure}

Furthermore, region identification alone is insufficient for reliable assertion generation. Assertions within each region must precisely map high-level functional behaviors to concrete RTL implementation. Errors in region identification or signal-role binding may propagate through subsequent behavior extraction and assertion generation, ultimately producing incorrect or incomplete SVAs. Therefore, effective CPU-level assertion generation requires a mechanism that bridges architectural semantics and low-level RTL structures, enabling reliable translation of functional intent into verification properties.

To address these challenges, we propose MicroAssert, a microarchitecture-aware assertion generation framework for CPU-level verification. MicroAssert identifies verification-relevant microarchitectural regions and generates assertions through behavior-specific primitives, where LLMs extract structured verification intents rather than directly generating SVAs. The instantiated primitives are then automatically parsed into deterministic SVA properties, avoiding unconstrained assertion generation and improving syntax correctness and generation reliability, as illustrated in Fig.~\ref{fig:motivation}.

The contributions of this paper are summarized as follows:

\begin{itemize}

    \item To the best of our knowledge, MicroAssert is the first framework that organizes CPU-level assertion generation around microarchitectural structures. By capturing architectural behaviors and cross-component interactions, MicroAssert enables accurate assertion generation and effective verification of complex CPU processors.

    \item We introduce behavior-specific primitives tailored to the functional characteristics of different microarchitectural regions. MicroAssert guides LLMs to fill these primitives with structured verification intents, which are automatically parsed into SVA properties, achieving 100\% syntax correctness and an average 74.82\% proven rate across CPU designs of varying scales and complexities.

    \item MicroAssert was evalutaed on five open-source RISC-V processors and achieve 100\% branch, statement, and toggle coverage. MicroAssert further achieves 100\% Cone of Influence coverage, an average 90.96\% Proof Core, and a significantly higher mutation-based error detection rate than existing approaches.

\end{itemize}

\section{Background}
\subsection{Assertion Generation Based on Traditional Methods}
Early assertion generation methods mainly relied on static
analysis and simulation traces. GoldMine~\cite{vasudevan2010goldmine}
combined data mining with static analysis to extract Boolean
and temporal properties from simulation traces.
Malburg et al.~\cite{malburg2017property} proposed a
dependency-graph-based approach that uses symbolic
abstraction and model checking to mine properties from
limited simulation data.
HARM~\cite{germiniani2022harm} further introduced
user-defined temporal logic templates and entropy-based
decision trees to guide assertion mining.

Although these methods automate assertion generation,
their dependence on predefined templates, static analysis,
or observed traces limits their ability to capture complex
functional behaviors and high-level design intent.
\subsection{Assertion Generation Based on LLMs}
Recent advances in large language models (LLMs) have
enabled assertion generation from natural language
specifications and RTL code.
Kande et al.~\cite{kande2024security} explored the use
of LLMs for generating hardware security assertions.
AssertLLM~\cite{fang2024assertllm} employed multiple
specialized LLMs to extract verification properties
from design specifications and generate SVAs.
AssertionForge~\cite{bai2025assertionforge}
constructed a unified knowledge graph from specifications
and RTL to connect high-level functional requirements
with implementation details.
More recently, DeepAssert~\cite{wang2025deepassert}
derived module-level specifications from RTL and used
them to generate fine-grained assertions for more
precise verification.

However, existing methods lack explicit microarchitectural identification and region localization, limiting their ability to systematically generate assertions for specific microarchitectural functions.

\section{MicroAssert Framework}

\subsection{Overview}

\begin{figure*}[t]
    \centering
    \includegraphics[width=0.95\textwidth]{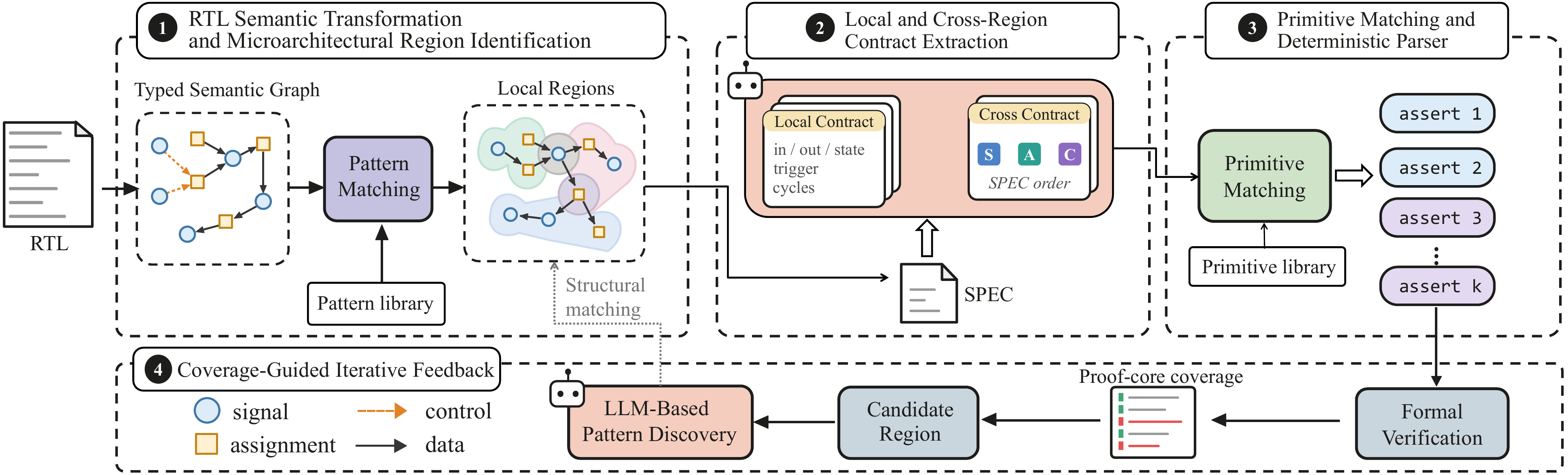}
    \caption{Overview of the MicroAssert framework.}
    \label{fig:framework}
\end{figure*}

CPU execution is organized around microarchitectural functions rather than RTL module boundaries. However, existing assertion generation approaches typically rely on RTL modules or code structures, overlooking that a single function may span multiple modules, while a single module may implement multiple functions. Consequently, module-centric analysis may fail to identify coherent functional regions and their behavioral dependencies, motivating a microarchitecture-aware approach that partitions RTL by functional. Guided by this idea, we propose MicroAssert, a microarchitecture-aware framework for structured assertion generation. As shown in Fig.~\ref{fig:framework}, MicroAssert consists of four stages: 1$)$ Microarchitectural Region Identification. 2$)$ Local and Cross-Region Contract Extraction. 3$)$ Primitive Matching and Deterministic Assertion Generation. 4$)$ Coverage-Guided Iterative Feedback.

\subsection{Microarchitectural Region Identification}

\begin{figure}[t]
    \centering
    \includegraphics[width=0.95\columnwidth]{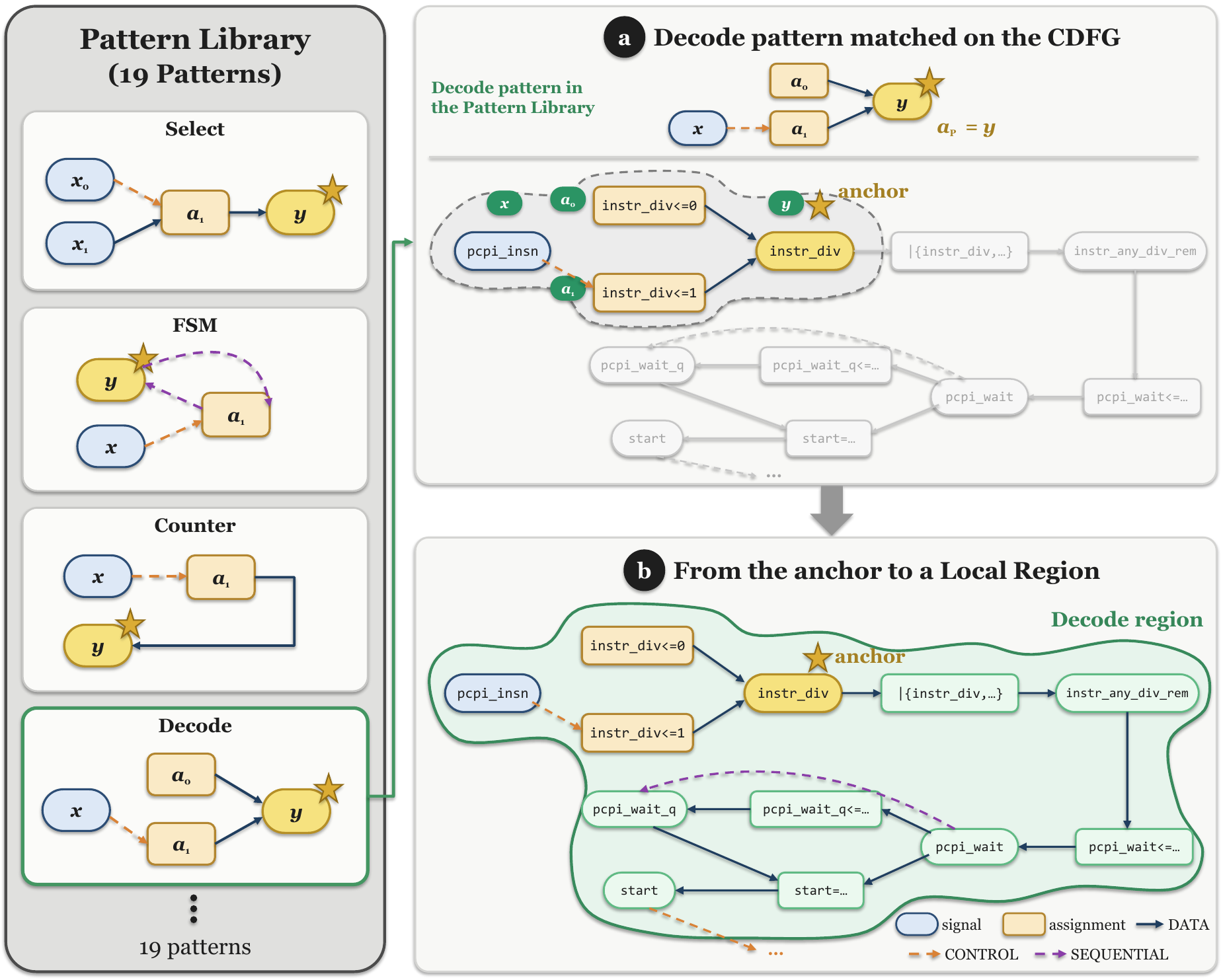}
    \caption{Example of identifying the \texttt{Decode} region from PicoRV32.}
    \label{fig:local_region}
\end{figure}

RTL code organization does not necessarily follow microarchitectural functional boundaries, as functionally related statements are often scattered across different assignments and conditional branches, making their intrinsic relationships difficult to identify from code locations alone. MicroAssert addresses this challenge by modeling microarchitectural functional region identification as a multi-relational graph pattern matching problem. First, MicroAssert converts RTL into a control and data flow graph (CDFG) $G = (V, E)$, where the node set $V$ comprises signal nodes and assignment nodes, and the directed edge set $E \subseteq V \times \mathcal{R} \times V$ explicitly captures data flow, control flow, and sequential transitions:
\[
\mathcal{R} = \{\mathrm{DATA},\; \mathrm{CTRL},\; \mathrm{SEQ}\},
\]
On this basis, we define each target microarchitectural structure as an anchored graph pattern:
\[
P = (V_P, E_P, a_P), \quad E_P \subseteq V_P \times \mathcal{R} \times V_P,
\]
where $V_P$ and $E_P$ specify the node attribute constraints and multi-relational dependency topology of the target microarchitecture, and $a_P \in V_P$ denotes the anchor signal used to locate the structure. Based on this formulation, we construct a fixed pattern library comprising 19 common microarchitectural structures (e.g., instruction decoding, state transitions, and pipeline control), as illustrated in Fig.~\ref{fig:local_region}.


To match a pattern $P$ within the global CDFG, MicroAssert treats all internally driven signals (i.e., signals written by at least one assignment in the RTL) as candidate anchors. For each candidate anchor $s$, MicroAssert extracts a local multi-relational candidate subgraph $G_s = (V_s, E_s)$ around its writing assignments, jointly collecting all three types of dependency edges:
\[
E_s = E_{\mathrm{DATA}}(s) \cup E_{\mathrm{CTRL}}(s) \cup E_{\mathrm{SEQ}}(s),
\]
where $E_{\mathrm{DATA}}(s)$ captures the assignment nodes writing to $s$ and the data signals read by their right-hand-side expressions, $E_{\mathrm{CTRL}}(s)$ captures the branch control conditions governing these assignments, and $E_{\mathrm{SEQ}}(s)$ captures sequential state transitions across register boundaries. Subsequently, MicroAssert aligns the candidate anchor $s$ with the pattern anchor $a_P$ and employs a VF2-based error-tolerant subgraph matching algorithm \cite{VF2} to compute the probabilistic structural similarity between $G_s$ and the pattern $P$, ensuring that target microarchitectural structures can be robustly identified even when different RTL implementations exhibit local topological deviations from the canonical pattern due to coding-style variations.

Since pattern matching focuses on core structural motifs, it may miss auxiliary assignments to matched signals. MicroAssert therefore collects all assignments to these signals across conditional branches and iteratively expands each matched subgraph along CDFG dependencies. Specifically, it traces upstream through $\mathrm{DATA}$ and $\mathrm{CTRL}$ edges to include producer assignments and branch conditions, downstream to capture dependent assignments and their target signals, and across register boundaries via $\mathrm{SEQ}$ edges to account for cross-cycle dependencies.

The resulting signals, assignments, and dependencies form a \emph{Local Region}, a pattern-anchored unit of logic analysis. MicroAssert records how internal logic reads external signals and how external logic uses internal signals, defining the region's input and output boundaries, respectively. Since expansion follows dependencies specific to each pattern match, signals and assignments may belong to multiple regions. Thus, Local Regions can overlap rather than form a mutually exclusive partition of the RTL code. A concrete example of identifying a \texttt{Decode} functional region is shown in Fig.~\ref{fig:local_region}.

\subsection{Local and Cross-Region Contract Extraction}

Identifying microarchitectural regions establishes the structural basis for verification, but does not specify their expected behaviors. MicroAssert therefore extracts behavioral requirements from the SPEC and associates them with corresponding Local Regions.

\textbf{Local Contract Extraction:} A Local Region identifies functionally related signals and assignments but has not yet linked them to the SPEC. MicroAssert establishes this link through a Local Contract, which records the region's behavioral rules, inputs, outputs, states, and trigger conditions. For multi-cycle behaviors, it explicitly extracts the cycle latency specified in the SPEC (e.g., single-cycle register update $+1\text{ cyc}$) between the trigger and expected outcome to ensure accurate assertion timing.

\begin{figure}[t]
    \centering
    \includegraphics[width=\linewidth]{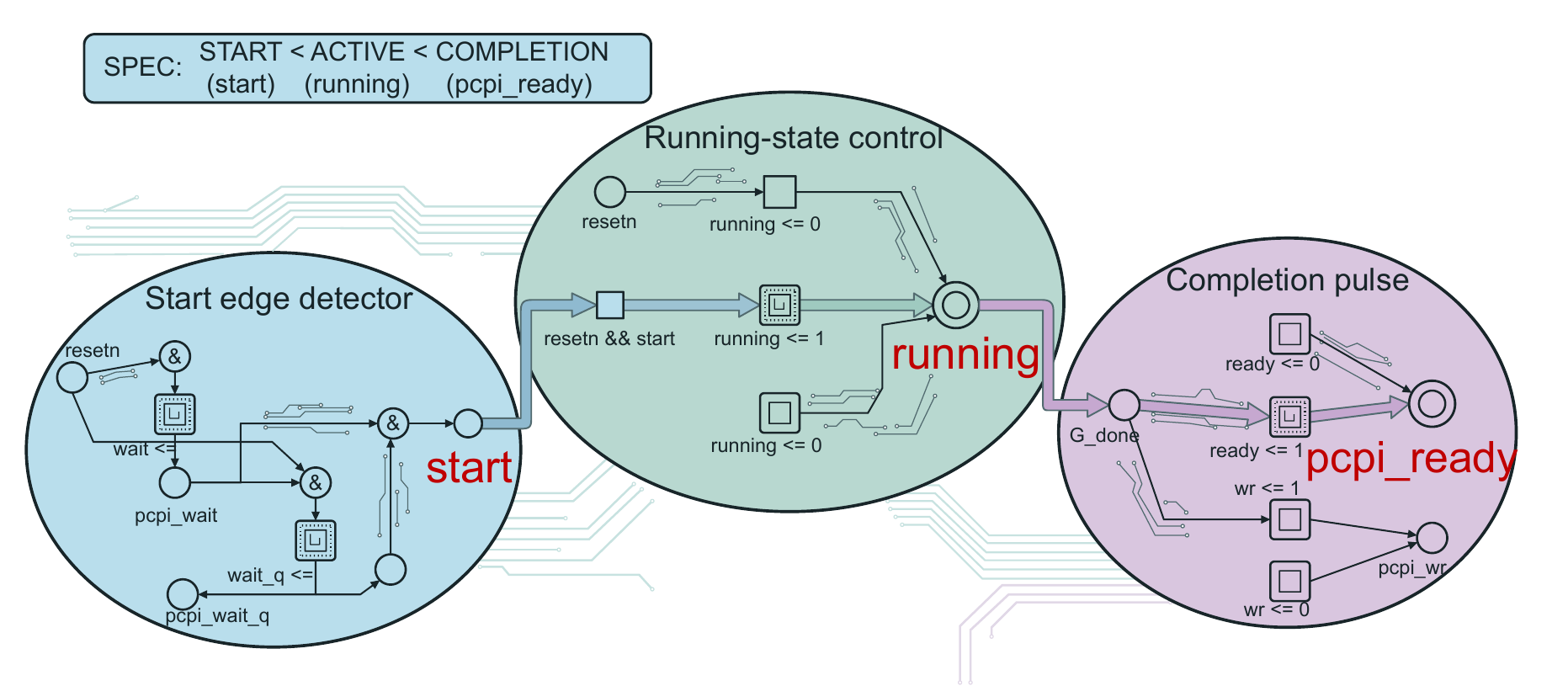}
    \caption{Example of cross-region event composition.}
    \label{fig:cross_region}
\end{figure}

\textbf{Cross-Region Contract Extraction:} Local Contracts capture individual region behaviors, whereas processor operations often span multiple regions. MicroAssert maps SPEC-defined execution relationships to events exposed by different Local Contracts, forming an event graph. It then connects compatible events into cross-region execution chains while preserving both qualitative stage ordering and quantitative latency bounds specified in the SPEC. As shown in Fig.~\ref{fig:cross_region}, the SPEC dictates that once triggered (\texttt{start}), the divider enters the active state (\texttt{running}) on the next cycle (\texttt{+1}) and remains active while shifting the 32-bit quotient mask until asserting \texttt{pcpi\_ready} within $M=34$ cycles (\texttt{RULE X1} in Fig.~\ref{fig:assert_case}).

MicroAssert retains only SPEC-supported relationships to avoid semantically irrelevant compositions.

\subsection{Primitive Matching and Deterministic Assertion Generation}
\label{sec:primitive}

Instead of directly generating assertions using an LLM, MicroAssert employs a predefined library of verification primitives covering common behavioral patterns, such as state updates, signal stability, and causal dependencies. The LLM is only responsible for mapping the extracted contracts to the corresponding primitives, while a parser translates them into syntactically correct assertions.

\textbf{Primitive Construction and Representativeness:} We derive 31 verification primitives by analyzing architectural specifications and RTL implementations across standard processor components. Each behavioral requirement is decomposed into a trigger condition, a target signal, an expected outcome, and a temporal relation. Design-specific signals, constants, and expressions are parameterized, while requirements sharing the same temporal-logical structure are consolidated into SVA templates with explicit timing and binding constraints.

Although processor microarchitectures vary widely, their verification properties are governed by five orthogonal behavioral dimensions, represented by the five core primitives in Table~\ref{tab:matching_patterns}. For example, \texttt{UPDATE} captures synchronous state and control updates, specifying how a target signal should change in the next cycle under a given condition. The remaining 26 primitives extend these foundational forms to accommodate more specific temporal and functional requirements, enabling the library to represent diverse CPU microarchitectural behaviors.

\begin{figure}[t]
    \centering
    \includegraphics[width=\linewidth]{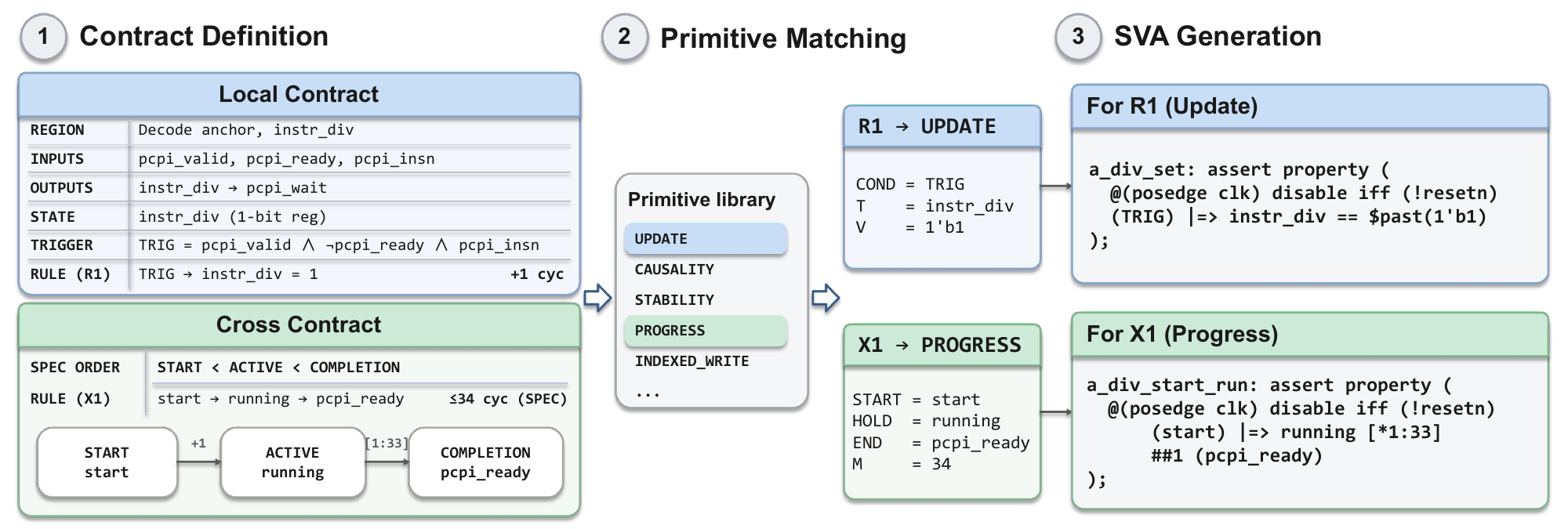}
    \caption{Local and cross-region contract-based SVA generation based on Fig.~\ref{fig:local_region} and Fig.~\ref{fig:cross_region}.}
    \label{fig:assert_case}
\end{figure}

\textbf{Primitive Matching and Deterministic Parsing:} For each extracted Local and Cross-Region Contract, MicroAssert maps its behavioral rules to corresponding verification primitives and binds the required signals, conditions, and temporal parameters to template roles. Specifically, deterministic matching rules jointly inspect the contract's SPEC-derived temporal semantics (e.g., single-cycle causal implication vs. bounded multi-cycle progress with deadline $M$) and the underlying CDFG attributes---including assignment types, branch conditions, signal widths, and data dependencies. For example, when a local contract rule (\texttt{R1} in Fig.~\ref{fig:assert_case}) specifies that a clocked assignment sets a target register under a trigger condition (\texttt{+1 cyc}), MicroAssert selects the \texttt{UPDATE} primitive, binding the trigger condition to \texttt{COND}, the assigned signal to \texttt{T}, and the right-hand-side expression to \texttt{V}. Similarly, for the cross-region contract rule \texttt{X1} shown in Fig.~\ref{fig:cross_region}, MicroAssert selects the \texttt{PROGRESS} primitive and binds the boundary and intermediate roles (\texttt{START = start}, \texttt{HOLD = running}, \texttt{END = pcpi\_ready}) along with the SPEC-derived latency bound ($M = 34$), compiling them into the multi-cycle property \texttt{(start) |=> running [*1:33] \#\#1 (pcpi\_ready)} shown in Fig.~\ref{fig:assert_case}.
MicroAssert then invokes a deterministic parser to instantiate the matched primitives into SVA property bodies and assemble complete assertions with clocking and reset conditions, guaranteeing syntactic correctness.

\begin{table}[t]
\centering
\caption{Representative behavior-specific SVA primitives in MicroAssert.}
\label{tab:matching_patterns}
\scriptsize
\setlength{\tabcolsep}{2.5pt}
\renewcommand{\arraystretch}{1.05}
\begin{tabularx}{\columnwidth}{
@{}
>{\raggedright\arraybackslash}p{0.24\columnwidth}
>{\raggedright\arraybackslash}X
@{}
}
\toprule
\textbf{Primitive} & \textbf{SVA Property Body} \\
\midrule
UPDATE &
\texttt{(COND) |=> T == \$past(V)} \\
CAUSALITY &
\texttt{(CAUSE) |-> (EFFECT)} \\
STABILITY &
\texttt{(COND) |=> \$stable(SIGS)} \\
PROGRESS &
\texttt{(START) |=> HOLD [*1:M-1] \#\#1 (END)} \\
INDEXED\_WRITE &
\texttt{(COND) |=> BANK[\$past(IDX)] == \$past(V)} \\
$\cdots$ & $\cdots$ \\
\bottomrule
\end{tabularx}
\end{table}


\subsection{Coverage-Guided Iterative Feedback}
\label{sec:feedback}

Successful formal proofs do not guarantee sufficient RTL coverage, since each proof may depend on only a subset of the associated logic, represented by its proof core. MicroAssert therefore guides iterative assertion generation using proof-core coverage. After each verification round, uncovered RTL statements reported by JasperGold are mapped back to CDFG nodes and edges. These uncovered nodes either reveal missing branch behaviors within existing Local Regions or indicate functional logic that bypassed initial graph pattern matching due to unconventional coding styles or severe topological variations. MicroAssert groups uncovered assignments by their target signals and traces their surrounding $\mathrm{DATA}$, $\mathrm{CTRL}$, and $\mathrm{SEQ}$ dependencies to construct candidate local subgraphs, and feeds the uncovered RTL code, the corresponding local CDFG subgraph, and its input/output boundary signals to the LLM to perform \emph{semantic structural matching}. 

Once the microarchitectural structure of a candidate region is resolved, it seamlessly re-enters MicroAssert's deterministic compilation pipeline. Finally, MicroAssert formally verifies the new assertions in JasperGold, remeasures cumulative proof-core coverage, and records newly covered statements and remaining coverage gaps to guide the next iteration until coverage converges.

\section{Expermential}

\subsection{Experimental Setup}
We evaluate MicroAssert on five open-source RISC-V CPU designs ~\cite{picorv32,biriscv, tinyriscv, ue_riscv, e203}with diverse scales and microarchitectural complexities. Table~\ref{tab:cpu_designs} summarizes characteristics of these designs, including RTL lines of code (LoC), the number of RTL modules (Mod.), and synthesized cell counts (Cells) obtained by Yosys. We compare MicroAssert with three existing automatic assertion generation approaches: AssertMiner-pro (AM-Pro)~\cite{assertminerpro}, AssertionForge (AF)~\cite{assertionforge} and SANGAM (SA)~\cite{sangam}. All generated SystemVerilog Assertions are formally verified using Cadence JasperGold (v21.12.002)~\cite{jaspergold}, where verification coverage is measured to evaluate assertion effectiveness. MicroAssert iteratively generates assertions using Proof Core Coverage as feedback. The process terminates when the cumulative Proof Core Coverage remains unchanged between two consecutive iterations. Experiments are conducted on a server equipped with an Intel(R) Xeon(R) Gold 6148 CPU at 2.40~GHz and 629~GB RAM. All methods use Qwen3.8 as the underlying base model.

\begin{table}[t]
\centering
\caption{Summary of the evaluated RISC-V CPU designs.}
\label{tab:cpu_designs}
\scriptsize
\setlength{\tabcolsep}{6pt}
\begin{tabular}{llcl}
\toprule
\textbf{Design} &
\multicolumn{1}{l}{\textbf{LoC}} &
\multicolumn{1}{l}{\textbf{Mod.}} &
\multicolumn{1}{l}{\textbf{Cells}} \\
\midrule
PicoRV32              & 3,049  & 8  & 19,358 \\
TinyRISC-V            & 6,004  & 31 & 24,117 \\
UltraEmbedded RISC-V  & 6,833  & 19 & 30,381 \\
UltraEmbedded biRISC-V& 8,432  & 23 & 71,731 \\
E203                  & 20,848 & 67 & 3,324,329 \\
\bottomrule
\end{tabular}
\end{table}

\subsection{Evaluation Metrics}

We evaluate assertion generation from three perspectives: generation reliability, design coverage, and verification effectiveness. Table~\ref{tab:metrics} summarizes the evaluation metrics, these are assessed using generation and proof statistics, formal coverage metrics, and Cone of Influence (COI)/Proof Core (PC) analysis with mutation testing, respectively. All generated assertions are evaluated without manual filtering or post-processing. 

\begin{table}[t]
\centering
\caption{Summary of evaluation metrics.}
\label{tab:metrics}
\scriptsize
\setlength{\tabcolsep}{6pt}
\renewcommand{\arraystretch}{1}
\begin{tabularx}{\columnwidth}{
@{}
>{\centering\arraybackslash}p{0.18\columnwidth}
>{\raggedright\arraybackslash}X
@{}
}
\toprule
\textbf{Metric} & \textbf{Summary} \\
\midrule
$N$   & Number of generated SVAs \\
SCR   & Rate of syntax-correct SVAs \\
PR    & Rate of proven SVAs among generated SVAs \\
SC    & Statement coverage in formal analysis \\
BC    & Branch coverage in formal analysis \\
TC    & Toggle coverage in formal analysis \\
COI   & Design logic potentially influencing a property \\
PC    & Design logic required to prove a property \\
ER    & Error detection rate in mutation testing \\
\bottomrule
\end{tabularx}
\end{table}

\begingroup
\setlength{\abovecaptionskip}{3pt}
\setlength{\belowcaptionskip}{3pt}
\setlength{\tabcolsep}{6pt}
\renewcommand{\arraystretch}{1}

\setlength{\aboverulesep}{2pt}
\setlength{\belowrulesep}{2pt}

\begin{table*}[h]
\centering
\scriptsize
\caption{Evaluation of assertion generation reliability.}
\label{tab:generation_quality}

\resizebox{0.8\textwidth}{!}{%
\begin{tabular}{lcccccccccccc}
\toprule
\multirow{2}{*}{\textbf{Design}} &
\multicolumn{4}{c}{\textbf{$N$}} &
\multicolumn{4}{c}{\textbf{SCR (\%)}} &
\multicolumn{4}{c}{\textbf{PR (\%)}} \\
\cmidrule(lr){2-5}
\cmidrule(lr){6-9}
\cmidrule(lr){10-13}
&
Ours & AM-Pro & AF & SA &
Ours & AM-Pro & AF & SA &
Ours & AM-Pro & AF & SA \\
\midrule
PicoRV32
& 367 & 43 & 660 & 243
& \mastcell{100} & 90.70 & 52.88 & 36.21
& \mastcell{79.29} & 55.81 & 36.06 & 42.17 \\

TinyRISCV
& 218 & 167 & 348 & 82
& \mastcell{100} & 77.84 & 82.47 & 100.00
& \mastcell{83.94} & 72.46 & 33.05 & 52.16 \\

UE RISC-V
& 288 & 272 & 592 & 89
& \mastcell{100} & 86.03 & 80.74 & 98.88
& \mastcell{72.91} & 71.32 & 38.85 & 59.27 \\

biRISC-V
& 309 & 296 & 437 & 104
& \mastcell{100} & 86.82 & 74.83 & 100.00
& \mastcell{70.87} & 64.19 & 38.22 & 46.33 \\

E203
& 447 & 612 & 268 & 451
& \mastcell{100} & 83.82 & 69.78 & 100.00
& \mastcell{67.11} & 53.16 & 24.25 & 28.19 \\

\midrule
\textbf{Avg.}
& 325.8 & 278.0 & 461.0 & 193.8
& \mastcell{100} & 85.04 & 72.14 & 87.02
& \mastcell{74.82} & 63.39 & 34.09 & 45.62 \\
\bottomrule
\end{tabular}%
}

\end{table*}
\endgroup

\begingroup
\setlength{\abovecaptionskip}{3pt}
\setlength{\belowcaptionskip}{3pt}
\setlength{\tabcolsep}{3pt}
\renewcommand{\arraystretch}{1.05}

\setlength{\aboverulesep}{3pt}
\setlength{\belowrulesep}{3pt}

\begin{table*}[t]
\centering
\scriptsize
\caption{Comparison of formal verification coverage
and proof-relevant logic across different RISC-V
CPU designs.}
\label{tab:verification_results}

\resizebox{1\textwidth}{!}{%
\begin{tabular}{lcccccccccccccccccccc}
\toprule
\multirow{2}{*}{\textbf{Design}} &
\multicolumn{4}{c}{\textbf{BC (\%)}} &
\multicolumn{4}{c}{\textbf{SC (\%)}} &
\multicolumn{4}{c}{\textbf{TC (\%)}} &
\multicolumn{4}{c}{\textbf{COI (\%)}} &
\multicolumn{4}{c}{\textbf{PC (\%)}} \\

\cmidrule(lr){2-5}
\cmidrule(lr){6-9}
\cmidrule(lr){10-13}
\cmidrule(lr){14-17}
\cmidrule(lr){18-21}

&
Ours & AM-Pro & AF & SA &
Ours & AM-Pro & AF & SA &
Ours & AM-Pro & AF & SA &
Ours & AM-Pro & AF & SA &
Ours & AM-Pro & AF & SA \\
\midrule

PicoRV32
& \mastcell{100} & 74.81 & 74.81 & 74.81
& \mastcell{100} & 63.98 & 63.98 & 63.98
& \mastcell{100} & 93.65 & 93.65 & 93.65
& \mastcell{100} & 53.84 & 66.96 & 65.19
& \mastcell{99.93} & 42.72 & 57.19 & 54.12 \\

TinyRISCV
& \mastcell{100} & 94.48 & 94.48 & 94.48
& \mastcell{100} & 91.44 & 91.44 & 91.44
& \mastcell{100} & 85.55 & 85.55 & 85.55
& \mastcell{100} & 91.89 & 88.27 & 86.06
& \mastcell{99.96} & 87.38 & 51.60 & 26.29 \\

UE RISC-V
& \mastcell{100} & 82.33 & 82.33 & 82.33
& \mastcell{100} & 76.50 & 76.50 & 76.50
& \mastcell{100} & 93.59 & 93.59 & 93.59
& \mastcell{100} & 83.04 & 78.69 & 77.98
& \mastcell{92.94} & 78.72 & 67.02 & 35.21 \\

biRISC-V
& \mastcell{100} & 80.62 & 73.48 & 86.79
& \mastcell{100} & 76.71 & 71.87 & 80.63
& \mastcell{100} & 83.89 & 69.28 & 94.61
& \mastcell{100} & 81.16 & 82.71 & 80.80
& \mastcell{89.45} & 73.03 & 56.09 & 44.94 \\

E203
& \mastcell{100} & 75.87 & 75.87 & 75.47
& \mastcell{100} & 78.22 & 78.22 & 78.22
& \mastcell{100} & 72.11 & 71.67 & 74.15
& \mastcell{100} & 78.94 & 72.39 & 79.05
& \mastcell{72.54} & 43.87 & 38.53 & 17.62 \\

\midrule

\textbf{Avg.}
& \mastcell{100} & 81.62 & 80.19 & 82.78
& \mastcell{100} & 77.37 & 76.40 & 78.15
& \mastcell{100} & 85.76 & 82.75 & 88.31
& \mastcell{100} & 77.77 & 77.80 & 77.82
& \mastcell{90.96} & 65.14 & 54.09 & 35.64 \\

\bottomrule
\end{tabular}%
}
\end{table*}
\endgroup

\subsection{Evaluation of Generation Reliability}

Table~\ref{tab:generation_quality} summarizes the assertion generation results of four methods across five RISC-V CPU designs. MicroAssert is the only method achieving 100\% SCR across all evaluated designs, demonstrating the effectiveness of our primitive-based assertion generation. The consistent syntax correctness across diverse CPU designs highlights its reliability over direct LLM-based assertion generation. In addition, MicroAssert achieves the highest PR across all evaluated designs, with an average of 74.82\%, compared with 63.39\%, 34.09\%, and 45.62\% for AM-Pro, AF, and SA, respectively. This improvement is not achieved by generating more assertions. On most evaluated designs, MicroAssert generates fewer assertions than AF while still achieving higher PR. The advantage becomes more evident as design complexity increases. On E203, the largest CPU in our evaluation, the PR of AM-Pro and AF drops to 53.16\% and 24.25\%, respectively, while MicroAssert maintains 67.11\%. These results demonstrate that MicroAssert scales more effectively to complex CPU designs.

\subsection{Evaluation of Design Coverage}
We evaluate design coverage using SC, BC, and TC reported by Cadence JasperGold. As shown in Table~\ref{tab:verification_results}, MicroAssert achieves 100\% coverage across all three metrics on all five CPU designs, whereas all baselines achieve average coverage below 90\%. Although existing methods attain relatively high coverage on some smaller designs, their coverage remains incomplete and varies across metrics. The gap is particularly evident on E203, the largest design in our evaluation, where all baselines fall below 80\% across three metrics. In contrast, MicroAssert maintains complete coverage despite the increased design complexity. These results highlight the effectiveness of MicroAssert's microarchitecture-aware region identification and coverage-guided refinement in systematically covering RTL behaviors across CPUs of varying complexity.

\subsection{Evaluation of Verification Effectiveness}

We evaluate assertion effectiveness using COI, PC, and mutation testing (ER). As shown in Table~\ref{tab:verification_results}, MicroAssert achieves 100\% COI across all five designs and an average PC of 90.96\%, exceeding AM-Pro (65.14\%), AF (54.09\%), and SA (35.64\%). On the largest E203, all baselines remain below 80\% COI, while SA's PC drops to 17.62\%. These results demonstrate MicroAssert's ability to connect assertions with proof-relevant RTL logic across CPUs of varying complexity.

\begin{figure}[t]
    \centering
    \includegraphics[width=1\columnwidth]{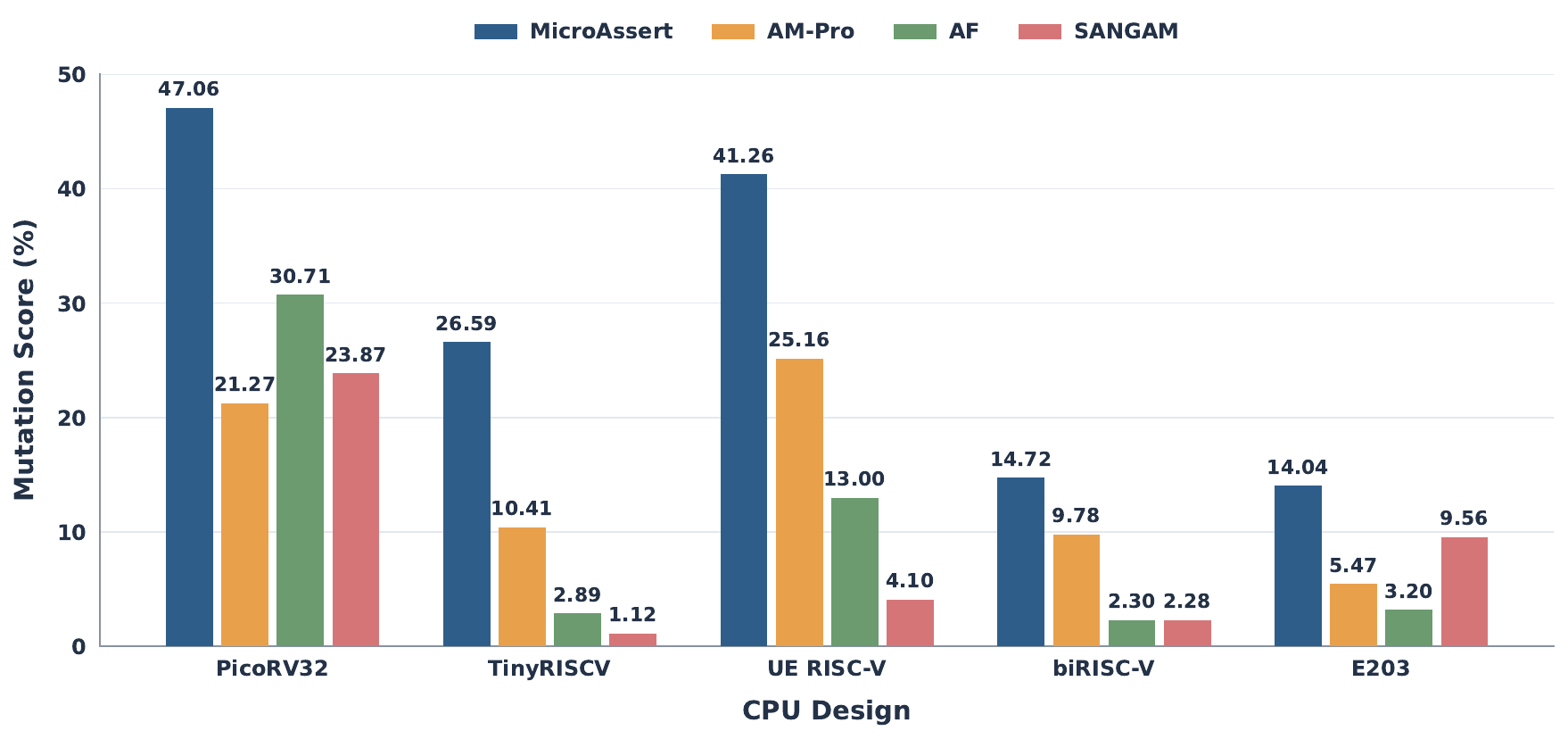}
    \caption{Mutation testing results across five RISC-V CPUs.}
    \label{fig:mutation_er}
\end{figure}

We use JasperGold’s built-in mutation testing flow to evaluate how effectively the generated assertions detect injected RTL errors. Fig.~\ref{fig:mutation_er} compares the ER reported by JasperGold for the four methods across all five CPU designs. MicroAssert achieves the highest ER on all five designs, averaging 28.73\%, which is 1.99$\times$, 2.76$\times$, and 3.51$\times$ the rates of AM-Pro (14.42\%), AF (10.42\%), and SA (8.19\%), respectively. Although the absolute ER may appear modest, mutation testing is highly stringent, as complex designs may involve millions of injected faults, and detecting subtle faults requires assertions to precisely capture their triggering conditions. Notably, MicroAssert achieves ERs of 47.06\% on PicoRV32 and 41.26\% on UE RISC-V. Since smaller CPUs contain fewer injected faults, the relative improvement in error detection is more pronounced on these designs. These results highlight MicroAssert's potential for effective error detection across processors of varying complexity.

\subsection{Ablation Study}

\subsubsection{Effectiveness of Architecture-Aware Region Identification}

To evaluate the effectiveness of our microarchitecture-aware region identification, we replace the proposed graph pattern matching and CDFG expansion method with direct LLM-based region identification, yielding a baseline variant (\texttt{w/o Arch-Aware}) for comparison. As shown in Table~\ref{tab:arch_aware_ablation}, MicroAssert consistently achieves substantially higher Proof Core (PC) coverage across all five CPUs while requiring significantly fewer assertions ($N$). Specifically, MicroAssert improves the average PC from 68.89\% to 90.96\% (+22.07 percentage points) while reducing the average number of generated assertions by 72.8\% (from 1,199.2 to 325.8). Without microarchitectural guidance, the LLM tends to fragment coherent hardware behaviors into isolated statement-level properties, producing excessive assertions with redundant or shallow proof cores. This fragmentation is particularly pronounced on complex designs such as the dual-issue biRISC-V, where \texttt{w/o Arch-Aware} generates 3,188 assertions ($10.3\times$ that of MicroAssert) yet still lags behind MicroAssert in PC by 12.68\% (76.77\% vs.\ 89.45\%). Similarly, on TinyRISCV and PicoRV32, MicroAssert reaches near-complete PC (99.96\% and 99.93\%) with roughly half the assertions, whereas \texttt{w/o Arch-Aware} plateaus at 62.42\% and 73.69\%, respectively.

\begin{table}[t]
\centering
\caption{Ablation study of architecture-aware assertion generation across five RISC-V CPU designs.}
\label{tab:arch_aware_ablation}
\scriptsize
\setlength{\tabcolsep}{3pt}
\renewcommand{\arraystretch}{1}
\begin{tabular}{lcccc}
\toprule
\multirow{2}{*}{\textbf{Design}}
& \multicolumn{2}{c}{\textbf{$N$}}
& \multicolumn{2}{c}{\textbf{PC (\%)}} \\
\cmidrule(lr){2-3}
\cmidrule(lr){4-5}
& Ours & w/o Arch-Aware
& Ours & w/o Arch-Aware \\
\midrule
PicoRV32  & 367 & 785  & \mastcell{99.93} & 73.69 \\
TinyRISCV & 218 & 407  & \mastcell{99.96} & 62.42 \\
UE RISC-V & 288 & 1037 & \mastcell{92.94} & 76.89 \\
biRISC-V  & 309 & 3188 & \mastcell{89.45} & 76.77 \\
E203      & 447 & 579  & \mastcell{72.54} & 54.68 \\
\midrule
\textbf{Avg.} & \textbf{325.8} & \textbf{1199.2} & \mastcell{\textbf{90.96}} & \textbf{68.89} \\
\bottomrule
\end{tabular}
\end{table}

A key reason for MicroAssert's superior coverage efficiency is its focus on intrinsic microarchitectural functionality rather than implementation-specific RTL module or code partitions. Different designers may organize the same processor functionality across varying numbers of RTL modules (e.g., PicoRV32 consolidates its core logic into 8 modules, whereas E203 distributes it across 67 modules). By capturing data, control, and sequential dependencies across module boundaries and grouping them into pattern-anchored Local Regions, MicroAssert moves beyond implementation-imposed divisions and concentrates assertion generation on complete, proof-relevant functional structures.

\subsubsection{Effect of Removing the Predefined SVA Primitives}

To evaluate the necessity of our verification primitive library and deterministic parser, we construct a variant (\texttt{w/o Prim.}) that prompts the LLM to directly synthesize raw SVA code from the extracted contracts. 

\begin{table}[h]
\centering
\caption{Ablation study of predefined SVA primitives (w/o Prim.: direct LLM assertion generation without SVA primitives).}
\label{tab:llm_wo_sva_template}
\scriptsize
\setlength{\tabcolsep}{5pt}
\renewcommand{\arraystretch}{1.15}
\begin{tabular}{lcccccc}
\toprule
\multirow{2}{*}{\textbf{Design}}
& \multicolumn{2}{c}{\textbf{SCR (\%)}}
& \multicolumn{2}{c}{\textbf{PR (\%)}} 
& \multicolumn{2}{c}{\textbf{PC (\%)}}\\
\cmidrule(lr){2-3}
\cmidrule(lr){4-5}
\cmidrule(lr){6-7}
& Ours & w/o Prim.
& Ours & w/o Prim.
& Ours & w/o Prim.\\
\midrule
PicoRV32  & \mastcell{100.00} & 97.04 & \mastcell{79.29} & 72.01 & \mastcell{99.93} & 93.75\\
TinyRISCV & \mastcell{100.00} & 96.92 & \mastcell{83.94} & 63.50 & \mastcell{99.96} & 97.43\\
UE RISC-V & \mastcell{100.00} & 97.23 & \mastcell{73.38} &72.91  & \mastcell{92.94} & 84.93\\
biRISC-V  & \mastcell{100.00} & 92.50 & \mastcell{70.87} & 58.80 & \mastcell{89.45} & 85.71\\
E203      & \mastcell{100.00} & 98.50 & \mastcell{67.11} & 66.62 & \mastcell{72.54} & 68.45\\
\midrule
\textbf{Avg.} & \mastcell{\textbf{100.00}} & \textbf{96.44} & \mastcell{\textbf{74.82}} & \textbf{66.86} & \mastcell{\textbf{90.96}} & \textbf{86.05}\\
\bottomrule
\end{tabular}
\end{table}

As shown in Table~\ref{tab:llm_wo_sva_template}, removing predefined primitives degrades performance across all three metrics, reducing average SCR from 100.00\% to 96.44\%, PR from 74.82\% to 66.86\% (-7.96\%), and PC from 90.96\% to 86.05\% (-4.91\%). Without the syntactic and temporal constraints of structural primitives, free-form LLM generation frequently introduces syntax errors, invalid clock/reset bindings, or mismatched cycle delays—most notably on TinyRISCV and biRISC-V, where PR drops by 20.44\% and 12.07\%, respectively. Furthermore, even when free-form generation yields a comparable PR (e.g., 73.38\% vs.\ 72.91\% on UE RISC-V), its PC drops by 8.01\% (84.93\% vs.\ 92.94\%), as unconstrained LLMs tend to emit trivial, easily provable combinational checks while failing to accurately formulate complex multi-cycle sequential assertions.

\subsubsection{Effectiveness of Coverage-Guided Feedback}

\begin{figure}[t]
    \centering
    \includegraphics[width=0.9\columnwidth]{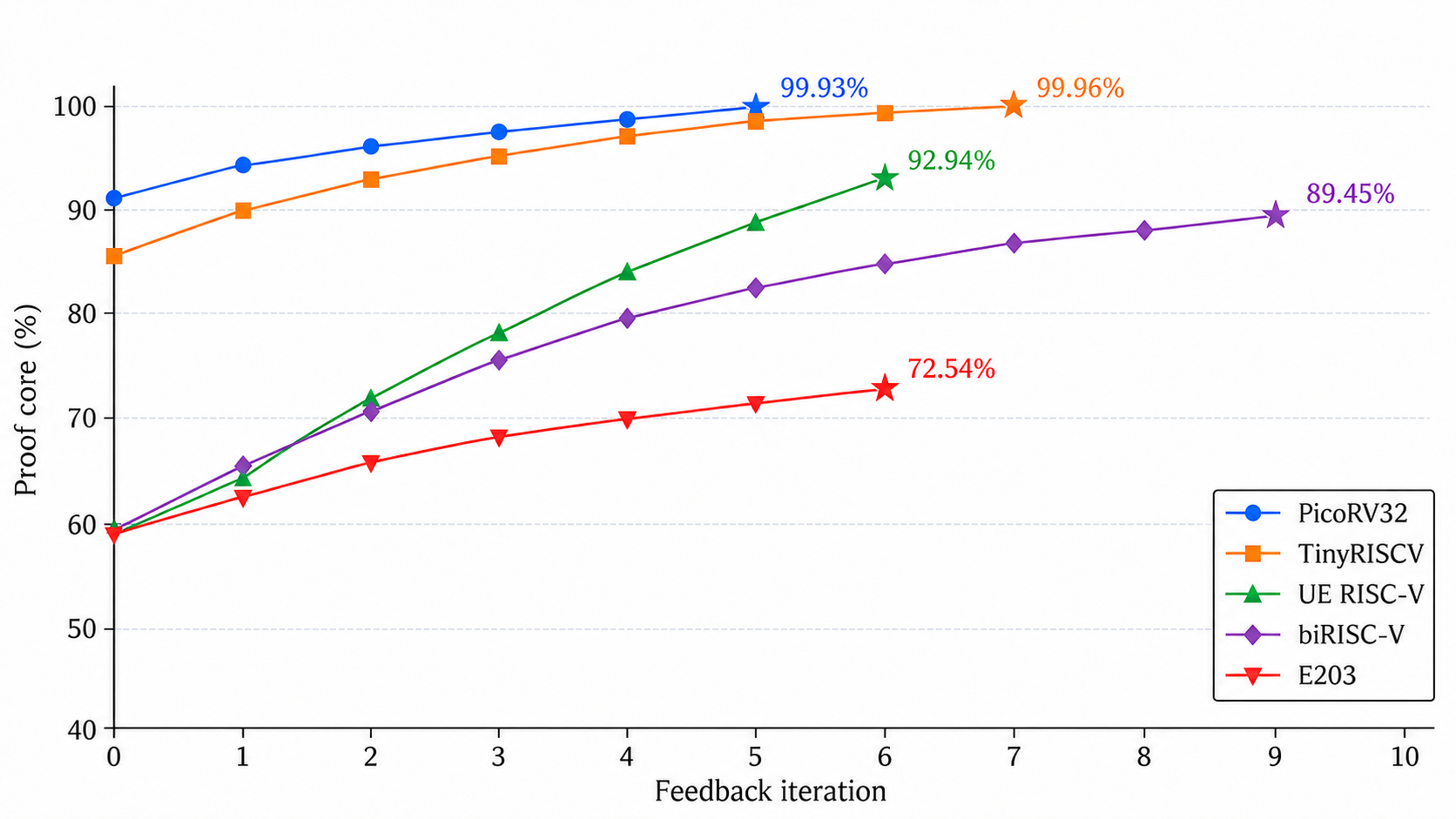}
    \caption{Proof Core evolution across ten feedback iterations.}
    \label{fig:iterative_feedback}
\end{figure}

Fig.~\ref{fig:iterative_feedback} tracks PC over ten feedback iterations, with iteration 0 representing the initial assertion set. PC reflects the RTL logic involved in assertion proofs and thus indicates whether identified regions capture meaningful behaviors. PC increases across all five CPUs, reaching 99.93\% and 99.96\% for PicoRV32 and TinyRISCV, 92.94\% and 89.45\% for UE RISC-V and biRISC-V, respectively. E203 reaches 72.54\% after six iterations. These results demonstrate that coverage-guided feedback progressively expands proof-relevant RTL coverage and refines the microarchitecture-level verification structure.

\subsubsection{Effect of Local and Cross-Region Assertions}

To evaluate the contribution of local and cross-region assertions, we construct two variants: w/o Local, which removes assertions derived from Local Contracts, and w/o Cross, which removes assertions derived from Cross-Region Contracts. The results are measured using Proof Core coverage.

\begin{table}[h]
\centering
\caption{Ablation study of local and cross-region assertions.}
\label{tab:local_cross_ablation}
\scriptsize
\setlength{\tabcolsep}{5pt}
\renewcommand{\arraystretch}{1.15}
\begin{tabular}{lccccc}
\toprule
& \multicolumn{2}{c}{\textbf{SVA Num.}}
& \multicolumn{3}{c}{\textbf{Ablation Results}}\\
\cmidrule(lr){2-3}
\cmidrule(lr){4-6}
\textbf{Design}
& \textbf{Local SVA}
& \textbf{Cross SVA}
& \textbf{Ours}
& \textbf{w/o Local}
& \textbf{w/o Cross}\\
\midrule
PicoRV32  & 238 & 129 & \mastcell{99.93} & 71.84 & 96.87\\
TinyRISCV & 125 &  93 & \mastcell{99.96} & 74.21 & 98.42\\
UE RISC-V & 224 &  64 & \mastcell{92.94} & 68.37 & 87.16\\
biRISC-V  & 232 &  77 & \mastcell{89.45} & 61.52 & 78.34\\
E203      & 283 & 164 & \mastcell{72.54} & 48.63 & 59.27\\
\midrule
\textbf{Avg.}
& \textbf{220.4}
& \textbf{105.4}
& \mastcell{\textbf{90.96}}
& \textbf{64.91}
& \textbf{84.01}\\
\bottomrule
\end{tabular}
\end{table}

As shown in Table~\ref{tab:local_cross_ablation}, removing local assertions substantially degrades PC across all five designs, reducing the average from 90.96\% to 64.91\%. In contrast, removing cross-region assertions has a smaller impact on smaller CPUs, but the degradation becomes more pronounced as design complexity increases, with PC dropping by 13.27\% on E203. These results demonstrate that local assertions provide fundamental region-level coverage, while the contribution of cross-region assertions becomes increasingly significant as CPU designs grow in scale and complexity, by capturing more complex interactions and dependencies across microarchitectural regions.

\section{Conclusion}

In this paper, we presented MicroAssert, a microarchitecture-aware framework for assertion generation in CPU verification. Through functional region identification, local and composite assertion generation, and coverage-guided refinement, MicroAssert captures microarchitectural behaviors and cross-region interactions. Experiments on five open-source RISC-V CPUs achieve 100\% syntax correctness and complete statement, branch, and toggle coverage, with average proven, Proof Core coverage, and mutation-based error detection rates of 74.82\%, 90.96\%, and 28.73\%, respectively. These results demonstrate the effectiveness of MicroAssert for CPU verification.

\bibliographystyle{IEEEtran}
\bibliography{refs}

\end{document}